\documentclass[aps,prb,citeautoscript,amsmath,amssymb,preprint,superscriptaddress]{revtex4-1}

\usepackage[utf8]{inputenc}
\usepackage[T1]{fontenc}
\usepackage{upgreek}
\usepackage[english]{babel}
\usepackage{amsmath}
\usepackage{amssymb}

\usepackage{amstext}
\usepackage{graphicx}
\usepackage{lmodern}
\usepackage{float}
\usepackage{siunitx}
\usepackage{color}
\usepackage{ulem}

\renewcommand{\bm}[1]{\boldsymbol{\mathbf{#1}}}
\definecolor{blue}{rgb}{0,0.286,0.333}
\definecolor{red}{rgb}{0.866,0.0392,0.207}

\usepackage[colorlinks=true,citecolor=blue,urlcolor=blue]{hyperref}

\begin{document}

\title{Spin-dependent (inverse) spin Hall effect in Co\textsubscript{60}Fe\textsubscript{20}B\textsubscript{20}}

\author{Joel Cramer}
\affiliation{Institute of Physics, Johannes Gutenberg-University Mainz, 55099 Mainz, Germany}
\affiliation{Graduate School of Excellence Materials Science in Mainz, 55128 Mainz, Germany}

\author{Andrew Ross}
\affiliation{Institute of Physics, Johannes Gutenberg-University Mainz, 55099 Mainz, Germany}
\affiliation{Graduate School of Excellence Materials Science in Mainz, 55128 Mainz, Germany}

\author{Samridh Jaiswal}
\affiliation{Institute of Physics, Johannes Gutenberg-University Mainz, 55099 Mainz, Germany}
\affiliation{Singulus Technologies AG, 63796 Kahl am Main, Germany}

\author{Lorenzo Baldrati}
\affiliation{Institute of Physics, Johannes Gutenberg-University Mainz, 55099 Mainz, Germany}

\author{Romain Lebrun}
\affiliation{Institute of Physics, Johannes Gutenberg-University Mainz, 55099 Mainz, Germany}


\author{Mathias Kläui}
\email{Klaeui@uni-mainz.de}
\affiliation{Institute of Physics, Johannes Gutenberg-University Mainz, 55099 Mainz, Germany}
\affiliation{Graduate School of Excellence Materials Science in Mainz, 55128 Mainz, Germany}

\date{\today}

	\begin{abstract}
		
		In ferromagnetic metals, the interconversion of spin and charge currents via the spin Hall effect and its inverse can depend on the angle between the ferromagnet's magnetization and the spin current polarization direction.
		Here, such a spin-dependent (inverse) spin Hall effect is found in the ferromagnetic alloy Co\textsubscript{60}Fe\textsubscript{20}B\textsubscript{20}.
		In a non-local magnon transport experiment, Co\textsubscript{60}Fe\textsubscript{20}B\textsubscript{20} is used to either excite or detect magnonic spin currents flowing in the ferrimagnetic insulator Y\textsubscript{3}Fe\textsubscript{5}O\textsubscript{12}.
		We find that the signal amplitude is significantly modulated by tuning the direction of the Co\textsubscript{60}Fe\textsubscript{20}B\textsubscript{20} magnetization.
		We design a sample structure that prevents direct magnonic coupling between the ferromagnets.
		Thus, we can identify unambiguously an intrinsic electronic origin of the observed effect.

	\end{abstract}

\maketitle

Pure spin currents that exclusively carry angular momentum have become a key feature of various spintronics concepts as they allow one to transport\cite{Chumak2015,Baltz2018} or manipulate\cite{Bhatti2017} information in magnetic systems in the absence of net charge motion.
In general, pure spin currents can be carried by an equal number of electrons with opposite spin momenta moving in opposite directions or magnons, the quanta of spin waves\cite{bauer2011spin}.
Feasible methods to generate or detect pure spin currents are provided by the spin Hall effect (SHE) and its inverse (ISHE)\cite{Sinova2015}.
In conductors with strong spin orbit interaction, charge currents are converted into transverse electronic spin currents by means of the SHE, while the ISHE describes the reversed process.
When in contact with a magnetic insulator, this can be used, for instance, to either generate\cite{Cornelissen2015,lebrun2018tunable} or detect\cite{Saitoh2006} magnonic spin flow in the insulating magnetic layer.

So far, research mainly focused on spin Hall studies in non-magnetic materials such as 4\textit{d}/5\textit{d} transition metals and their alloys\cite{Sinova2015,Cramer2018a}.
Recently, metallic ferromagnets (FM) have also been shown to exhibit the (I)SHE \cite{Miao2013}.
However, in ferromagnets there is with the magnetization direction an additional degree of freedom, which could impact the conversion efficiency.
In the literature, contradicting claims have been put forward with no effect of the magnetization direction reported in Ref.~\onlinecite{Tian2016}, while for a fixed spin current polarization direction the detection efficiency was shown in Ref.~\onlinecite{Cramer2018,Das2017} to depend on the FM's magnetization orientation.
This has been used, for instance, in recent devices to implement a spin valve like effect\cite{Cramer2018}.
However, in both these previous experiments the FM detectors directly interface the magnetic system mediating the magnonic spin currents such that the latter couple to the FM ISHE detector.
Assuming a two spin current model with an effective interplay between electronic and magnonic spin currents in the FM due to \textit{sd}-exchange coupling\cite{Cheng2017}, the transferred angular momentum thus may be transported by both magnonic\cite{Barna1992,Cramer2018b} and electronic spin currents in the FM detector, potentially resulting in different effects.
Therefore, to unambiguously identify the effect of charge based spin currents for the detection efficiency via the ISHE  in a metallic ferromagnet, one needs to design a system where the magnonic contribution is suppressed.

In this work, the magnetization orientation-dependent interconversion of spin and charge information (i.e. SHE and ISHE) is probed in the metallic ferromagnet Co\textsubscript{60}Fe\textsubscript{20}B\textsubscript{20} (CoFeB).
Performing a non-local magnon transport experiment\cite{Cornelissen2015,Goennenwein2015, lebrun2018tunable}, the signal amplitudes induced by spin currents that are either triggered or detected in the CoFeB are measured as a function of the external field amplitude that sets the angle between the detector magnetization and the spin current polarization direction.
To exclude magnonic effects, we add a non-magnetic metal spacer layer, which exchange decouples the magnetic system that laterally transports the spin current (Y\textsubscript{3}Fe\textsubscript{5}O\textsubscript{12}) and the CoFeB injector/detector.
The resulting signal amplitudes are compared to magnetoresistance data, revealing a direct correlation between the spin signal amplitude and the direction of the CoFeB magnetization, demonstrating a spin-dependent (I)SHE.

\begin{figure}[!t]
	\centering
	\includegraphics[width=85 mm]{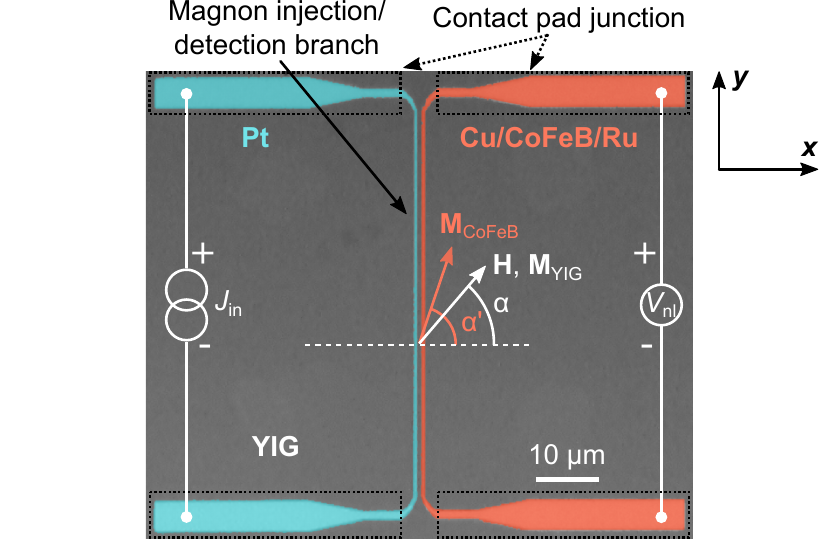}
	\caption{Colored optical micrograph of the implemented non-local device structure comprising two parallel, electrically insulated Pt (blue) and Cu/CoFeB/Ru (orange) wires.
		The electrical wiring as well as the magnetic field direction definition are indicated.
		The micrograph was taken before the deposition of contact pads.}
	\label{fig:cofeb1}
\end{figure}

Figure~\ref{fig:cofeb1} shows an optical micrograph of the implemented device structure, which includes two parallel, electrically insulated Pt and Cu/CoFeB/Ru wires deposited on the insulating ferrimagnet Y\textsubscript{3}Fe\textsubscript{5}O\textsubscript{12} (YIG).
The latter has a thickness of $d_\mathrm{YIG} = \SI{630}{\nano\meter}$ and is grown on a (111)-oriented Gd\textsubscript{3}Ga\textsubscript{5}O\textsubscript{12} substrate via liquid phase epitaxy.
The nanowires were produced by a multi-step lift-off procedure based on electron beam lithography.
First, nanowires with a width $w = \SI{250}{\nano\meter}$ were patterned, followed by the sputter deposition of \SI{7.5}{\nano\meter} of Pt.
For the second type of nanowire ($w = \SI{350}{\nano\meter}$, center-to-center distance to the first wire $d = \SI{1250}{\nano\meter}$), a Cu/CoFeB/Ru multilayer stack with individual layer thicknesses of $d_\mathrm{Cu} = \SI{2}{\nano\meter}$, $d_\mathrm{CoFeB} = \SI{7}{\nano\meter}$ and $d_\mathrm{Ru} = \SI{2}{\nano\meter}$ was deposited by magnetron sputtering after an additional lithography step.
While the Cu layer exchange decouples YIG and CoFeB, Ru acts as a capping layer.
For simplicity, the Cu/CoFeB/Ru multilayer is referred to as CoFeB in the following.

In the experiment, both the Pt and CoFeB wire are used for magnon injection and detection in order to study the charge-to-spin (SHE) as well as spin-to-charge (ISHE) conversion in CoFeB, respectively.
Applying a charge current to the injector results in both the thermal and electrical excitation of magnonic spin currents in YIG via, respectively, the spin Seebeck effect\cite{Uchida2010a} (SSE) and the SHE induced spin current in the injector\cite{Cornelissen2015}.
The magnons propagate in the magnetic insulator and, when absorbed by the detector, induce an electrical voltage response by means of the ISHE.
Implementing a DC measurement scheme, the injector charge current $J_\mathrm{in} = \pm\SI{300}{\micro\ampere}$ ($j_\mathrm{in}^\mathrm{Pt} \approx \SI{1.6E11}{\ampere\per\square\meter}$, $j_\mathrm{in}^\mathrm{CoFeB} \approx \SI{7.8E10}{\ampere\per\meter\squared}$) was supplied by a Keithley 2400 source meter, simultaneously recording the injector resistance.
As demonstrated later, magnetoresistance effects reveal relevant information on the magnetic configuration of the system.
At the detector, the non-local spin signal was picked up by a Keithley 2182A nanovoltmeter.
To separate electrically and thermally generated spin signals, the polarity of the charge current applied to the injector was reversed and the difference (electrically induced signal) or sum (thermally induced signal) of the corresponding non-local voltages was considered\cite{Goennenwein2015,Thiery2018}:
\begin{align}
V^\mathrm{nl}_\mathrm{\Delta} &= \left[ V_\mathrm{nl} \left( + J_\mathrm{in} \right) - V_\mathrm{nl} \left( - J_\mathrm{in} \right) \right] / 2., \\
V^\mathrm{nl}_\mathrm{\Sigma} &= \left[ V_\mathrm{nl} \left( + J_\mathrm{in} \right) + V_\mathrm{nl} \left( - J_\mathrm{in} \right) \right] / 2. 
\end{align}
For the spin transport measurements, an in-plane magnetic field $\bm{H}$ was applied and both angular-dependent (sample rotation in a fixed field, max. $\upmu_0 H = \SI{85}{\milli\tesla}$) and field sweep measurements at different angles (max. $\upmu_0 H = \SI{175}{\milli\tesla}$) were performed.
The field configuration is illustrated in Fig.~\ref{fig:cofeb1}, with the angle $\alpha\,=\,\ang{0}$ signifying an alignment of $\bm{H}$ in the positive $\mathbf{x}$-direction.
All measurements were conducted at room temperature.

\begin{figure}[!bt]
	\centering
	\includegraphics[width=85 mm]{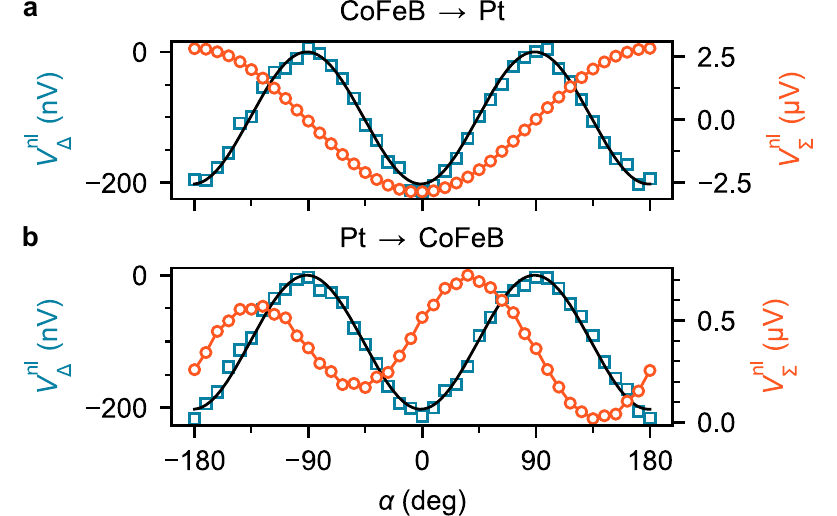}
	\caption{(a),(b) Angular-dependent non-local voltages induced by electrically ($V^\mathrm{nl}_\mathrm{\Delta}$) and thermally ($V^\mathrm{nl}_\mathrm{\Sigma}$) excited spin currents for a field larger than the saturation field.
		In (a) CoFeB is used as injector and Pt as detector, while results obtained for the reversed configuration are shown in (b). }
	\label{fig:cofeb2}
\end{figure}

\begin{figure}[!t]
	\centering
	\includegraphics[width=85 mm]{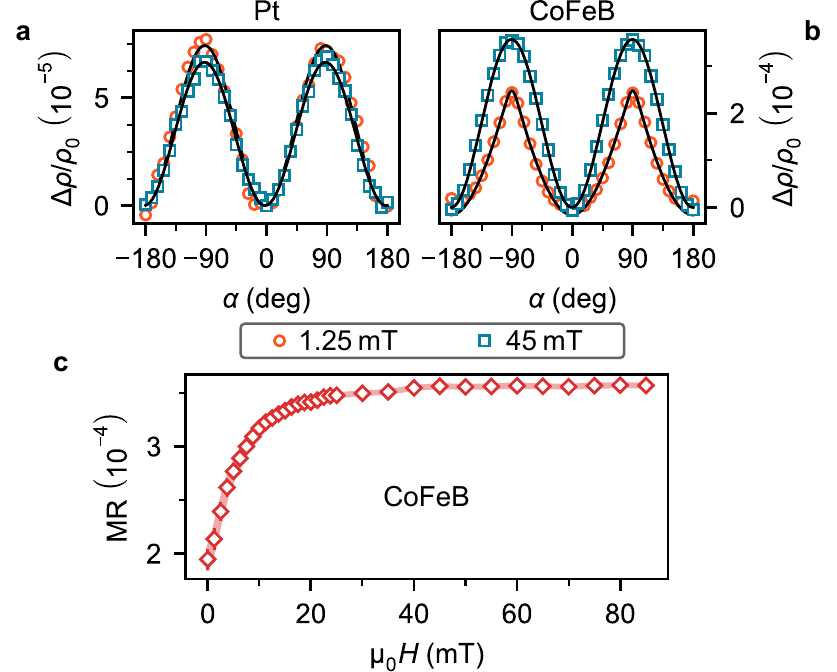}
	\caption{(a),(b) Angular-dependent magnetoresistance $\Delta\rho/\rho_0$ measured for (a) the Pt (SMR) and (b) the CoFeB wire (SMR + AMR, AMR dominating) for two different external magnetic fields.
		(c) Amplitude of the angular-dependent magnetoresistance (MR) in CoFeB as a function of the external magnetic field amplitude.
		Data at $\upmu_0 H = \SI{0}{\milli\tesla}$ are obtained with the system being in the remanent state.
		Error bars represent the (a), (b) standard error of the mean and (c) propagated errors.}
	\label{fig:cofeb3}
\end{figure}

First, the behavior of the non-local device was verified by angular-dependent measurements capturing both the electrically and thermally excited signal.
In Fig.~\ref{fig:cofeb2}a,b, $V^\mathrm{nl}_\mathrm{\Delta}$ (circles) and $V^\mathrm{nl}_\mathrm{\Sigma}$ (squares) are shown as a function of $\alpha$ for the two distinct configurations of either using CoFeB as injector and Pt as detector (CoFeB $\rightarrow$ Pt, Fig.~\ref{fig:cofeb2}a) or vice versa (Pt $\rightarrow$ CoFeB, Fig.~\ref{fig:cofeb2}b).
Note that for these measurements the applied field of $\upmu_0 H = \SI{85}{\milli\tesla}$ is larger than the saturation field of both YIG and CoFeB (see Fig.~\ref{fig:cofeb3} or Supporting Information) so that their respective magnetizations are always completely aligned along the field direction.
Considering first the electrically induced signal, $V^\mathrm{nl}_\mathrm{\Delta}$ reveals for both configurations a $\cos^2\left(\alpha\right)$ angular dependence, which is given by the combination of the individual symmetries of the magnon injection and detection processes via SHE and ISHE, respectively\cite{Cornelissen2015}.
The signals have a negative sign and, within the error, equal amplitude with $V^\mathrm{nl}_\mathrm{\Delta} \left( \mathrm{CoFeB} \rightarrow \mathrm{Pt} \right) = \SI[separate-uncertainty = true]{-203 \pm 4}{\nano\volt}$ and $V^\mathrm{nl}_\mathrm{\Delta} \left( \mathrm{Pt} \rightarrow \mathrm{CoFeB} \right) = \SI[separate-uncertainty = true]{-202 \pm 3}{\nano\volt}$.
The negative sign of $V^\mathrm{nl}_\mathrm{\Delta}$ signifies that the ISHE charge current in the detector and $J_\mathrm{in}$ applied to the injector flow in the same direction, meaning that the spin Hall angle of CoFeB and Pt have the same sign\cite{Goennenwein2015}.

Regarding the thermal response, $V^\mathrm{nl}_\mathrm{\Sigma}$ detected by the Pt detector (Fig.~\ref{fig:cofeb2}a) exhibits a $\cos\left(\alpha\right)$ symmetry, which is, in the case of YIG/Pt, characteristic for an SSE induced signal \cite{Uchida2016}.
In the $\mathrm{Pt} \rightarrow \mathrm{CoFeB}$ configuration, the angular dependence of the voltage measured by the CoFeB detector significantly differs (Fig.~\ref{fig:cofeb2}b), exhibiting approximately a $\sin\left( 2 \alpha \right)$ variation.
Apart from a SSE current flowing in the YIG \cite{Uchida2016}, the signal can comprise further magneto-thermal effects like the anomalous \cite{Nagaosa2010} or planar \cite{Avery2012} Nernst effect that may arise in the CoFeB. 
The observed $\sin\left( 2 \alpha \right)$ symmetry possibly suggests a dominating planar Nernst effect, which is a result of an in-plane temperature gradient and the spin-thermoelectric equivalent of the anisotropic magnetoresistance (AMR\cite{Mcguire1975})\cite{Avery2012}.
Altogether, this result demonstrates that $V^\mathrm{nl}_\mathrm{\Sigma}$, which mainly relates to thermal effects, is not suitable to quantify the spin-charge interconversion in CoFeB.
This measurement scheme will therefore not be considered in light of this aspect hereafter.
Instead, this work will focus on $V^\mathrm{nl}_\mathrm{\Delta}$ originating from electrically induced magnon transport and that is free from such artifacts.

The necessary information on the magnetic configuration of the system is provided by magnetoresistance (MR) effects.
While for YIG/Pt only the spin Hall magnetoresistance (SMR) is expected to appear \cite{Nakayama2013}, both SMR and AMR can be observed in YIG/CoFeB.
The SMR amplitude depends on the angle between the polarization of the SHE induced spin current in the Pt or CoFeB wire and the YIG magnetization \cite{Nakayama2013}, while the AMR is determined by the angle between $J_\mathrm{in}$ and the magnetization direction in CoFeB \cite{Mcguire1975}.
Figure~\ref{fig:cofeb3}a shows angular-dependent SMR data obtained for the Pt injector at two distinct external magnetic fields $\upmu_0 H = \SI{1.25}{\milli\tesla}$ and $\upmu_0 H = \SI{45}{\milli\tesla}$.
For both field values, the resistance change $\Delta\rho / \rho_0$ reveals the expected $\sin^2\left(\alpha\right)$ symmetry\cite{Nakayama2013} with comparable SMR amplitudes, signifying that the YIG magnetization follows the external field even at low field strengths.
In the case of the CoFeB wire, both shape and amplitude of the angular-dependent magnetoresistance vary significantly for the different field amplitudes applied, see Fig.~\ref{fig:cofeb3}b.
At low fields, $\Delta\rho / \rho_0$ exhibits a flat resistance change near $\alpha = \ang{0},\pm\ang{180}$ and distinct peaks at $\alpha = \pm\ang{90}$, whereas the $\sin^2\left(\alpha\right)$ dependence characteristic for both SMR and AMR \cite{Mcguire1975} is seen at larger external fields.
Considering that the measured resistance is dominated by the magnon injection/detection branch of the wire, see Fig.~\ref{fig:cofeb1}, this behavior can be explained by a dominating AMR and a uniaxial shape anisotropy.
The latter forces $\bm{M}_\mathrm{CoFeB}$ to align along the wire axis at low fields, while at larger fields the Zeeman energy becomes dominating and $\bm{M}_\mathrm{CoFeB}$ aligns parallel to $\bm{H}$, irrespective of the field direction.

Bearing in mind that the direction of $\bm{M}_\mathrm{YIG}$ is fixed by the external field for basically all field values, the anisotropic MR recorded for CoFeB signifies that $\bm{M}_\mathrm{YIG}$ and $\bm{M}_\mathrm{CoFeB}$ are efficiently exchange-decoupled.
Furthermore, large external fields are required to achieve parallel alignment of $\bm{M}_\mathrm{YIG}$ and $\bm{M}_\mathrm{CoFeB}$ for external field directions that are perpendicular to the wire, which confirms that the studied system allows one to investigate the magnetization orientation-dependent generation/detection of spin currents in CoFeB.
To derive the angle $\phi = \alpha ' - \alpha$ between $\bm{M}_\mathrm{YIG}$ and $\bm{M}_\mathrm{CoFeB}$, see Fig.~\ref{fig:cofeb1}, one can principally solve the Stoner-Wohlfarth model \cite{coey2010magnetism} and reconstruct an angular-dependent magnetoresistance, as demonstrated by the solid line in Fig.~\ref{fig:cofeb2}b.
However, the additional MR in the nanowire contact pad junctions prevents an unambiguous quantification of $\phi$, such that the amplitude of the total MR effect in CoFeB $\Delta\rho / \rho_0 = \left[ \rho\left(\ang{90}\right)-\rho\left(\ang{0}\right)\right] / \rho\left(\ang{0}\right)$ is used to obtain qualitative information on the CoFeB magnetization orientation at different field strengths.
As shown in Fig.~\ref{fig:cofeb3}c, $\Delta\rho /\rho_0$ initially increases with increasing field amplitude until it saturates above a critical value of $\upmu_0 H_\mathrm{c} \geq \SI{25}{\milli\tesla}$.
Hence, $\bm{M}_\mathrm{CoFeB}$ is fully aligned with the external field ($\bm{M}_\mathrm{YIG}$) in all directions for $H \geq H_\mathrm{c}$, whereas a finite angle between $\bm{M}_\mathrm{YIG}$ and $\bm{M}_\mathrm{CoFeB}$ appears for small fields aligned perpendicular to the wire axis.

\begin{figure}[!t]
	\centering
	\includegraphics[width = 85 mm]{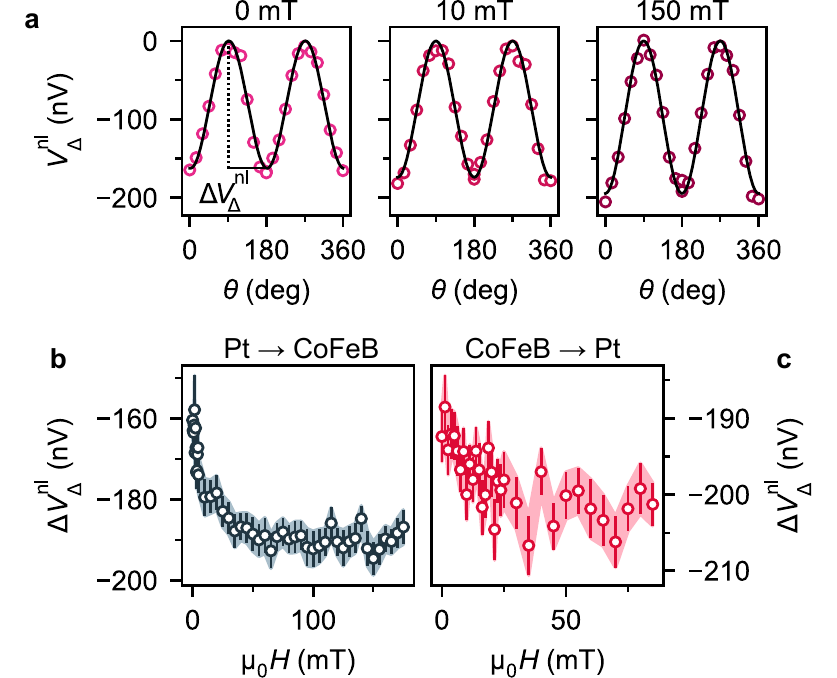}
	\caption{(a) Angular-dependent variation of $V_{\Delta}^\mathrm{nl}$ at different magnetic fields obtained in the Pt $\rightarrow$ CoFeB configuration.
		The data is extracted from uniaxial hysteresis loops at fixed angles between the external field and the wire axis ($\alpha = \ang{0}$: field perpendicular to the wire), with $\upmu_0 H = \SI{0}{\milli\tesla}$ implying remanence of system.
		Solid lines correspond to fit functions.
		(b),(c) Amplitude $\Delta V_{\Delta}^\mathrm{nl}$ of the angular-dependent voltage as a function of field for (b) the Pt $\rightarrow$ CoFeB configuration and (c) the CoFeB $\rightarrow$ Pt configuration.
		Error bars account for (a) the standard error and (b),(c) fit parameter errors.}
	\label{fig:cofeb5}
\end{figure}

The data shown so far demonstrates that the implemented non-local device indeed enables the electrical generation and detection of spin currents in CoFeB for zero (high field) and finite (low field) angles between the YIG and CoFeB magnetization.
As a result, one can probe both a potential spin orientation-dependent SHE and ISHE in the metallic ferromagnet.
Figure~\ref{fig:cofeb5}a shows $V_{\Delta}^\mathrm{nl}$ as a function of $\alpha$ and field amplitude in the Pt~$\rightarrow$~CoFeB configuration, in which CoFeB acts as the spin current detector.
To exploit the whole spectrum of available field amplitudes ($\SI{0}{\milli\tesla} \leq \upmu_0 H \leq \SI{175}{\milli\tesla}$), the angular-dependent data was extracted from uniaxial hysteresis loops measured at different angles $\alpha$ ($\upmu_0 H = \SI{0}{\milli\tesla}$ signifies the remanent state of the system, see Supporting Information).
As can be seen in the graphs, the amplitude $\Delta V_{\Delta}^\mathrm{nl}$ of the non-local voltage increases with increasing field and fitting yields $\Delta V_{\Delta}^\mathrm{nl} \left(\SI{0}{\milli\tesla}\right) = \SI[separate-uncertainty=true]{-162 \pm 4}{\nano\volt}$, $\Delta V_{\Delta}^\mathrm{nl} \left(\SI{10}{\milli\tesla}\right) = \SI[separate-uncertainty=true]{-173 \pm 5}{\nano\volt}$ and $\Delta V_{\Delta}^\mathrm{nl} \left(\SI{150}{\milli\tesla}\right) = \SI[separate-uncertainty=true]{-194 \pm 4}{\nano\volt}$.
The detailed field-dependence is shown in Fig.~\ref{fig:cofeb5}b, revealing an initial increase of the absolute signal amplitude, followed by a saturation.
The relative amplitude change between the low and high field level is $\left[ \Delta V_{\Delta}^\mathrm{nl} \left( >\SI{100}{\milli\tesla} \right)- \Delta V_{\Delta}^\mathrm{nl} \left(\SI{0}{\milli\tesla}\right)\right]/\Delta V_{\Delta}^\mathrm{nl} \left(\SI{0}{\milli\tesla}\right) = \SI[separate-uncertainty=true]{17.2 \pm 3.4}{\percent}$.
In the case of the reverse setup, i.e., CoFeB now acts as spin current injector and Pt as detector, a qualitatively similar but weaker effect amplitude of \SI[separate-uncertainty=true]{5.3 \pm 2.6}{\percent} is observed, see Fig.~\ref{fig:cofeb5}c.

\begin{figure}[!b]
	\centering
	\includegraphics[width=85 mm]{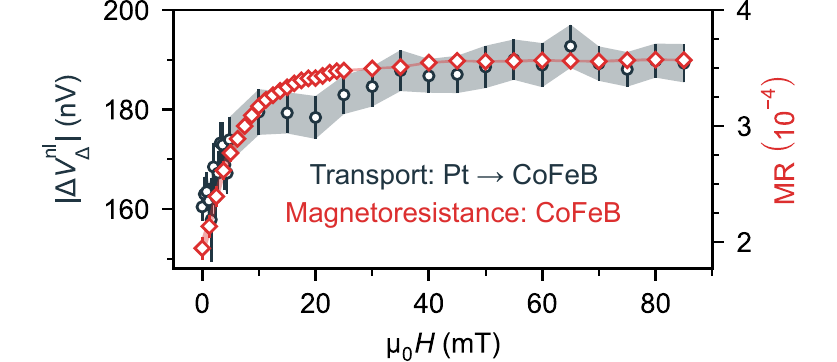}
	\caption{Direct comparison of $\Delta\rho / \rho_0$ (Fig.~\ref{fig:cofeb3}) and the non-local signal amplitude $\delta V_{\Delta}^\mathrm{nl}$ (Fig.~\ref{fig:cofeb5}b) in the CoFeB wire as a function of field.
		Data at $\upmu_0 H = \SI{0}{\milli\tesla}$ are obtained with the system being in the remanent state.
		Error bars are given by the standard error ($\Delta\rho / \rho_0$) and by extracted fit parameter errors ($ \Delta V_{\Delta}^\mathrm{nl}$).}
	\label{fig:cofeb6}
\end{figure}

To check whether this field-dependent modulation of the electrically induced spin signal correlates with the CoFeB magnetization orientation and, hence, can be explained by a spin-dependent SHE or ISHE, the field dependence of $\Delta V_{\Delta}^\mathrm{nl}$ in the Pt $\rightarrow$ CoFeB configuration is compared to that of the MR amplitude in CoFeB in Fig.~\ref{fig:cofeb6}.
Evidently, $\Delta\rho / \rho_0$ and $\left| \Delta  V_{\Delta}^\mathrm{nl}\right|$ exhibit the same qualitative field dependence with an initial amplitude increase and a saturation above $\upmu_0 H_\mathrm{c} \gtrsim \SI{25}{\milli\tesla}$.
As mentioned above, the magnetoresistance in CoFeB is AMR-dominated and hence directly correlates with the magnetization ($\bm{M}_\mathrm{CoFeB}$) direction.
The matching field dependences thus demonstrate that the spin-to-charge conversion efficiency in CoFeB exhibits a magnetization orientation dependence, which in this work is termed spin-dependent inverse spin Hall effect (see below).
Bearing in mind Onsager's reciprocity relations \cite{Onsager1931}, the same conclusion can be drawn for the SHE (see Fig.~\ref{fig:cofeb5}c).

The observation of a spin-dependent SHE and ISHE in CoFeB agrees with the findings reported by Das \textit{et al.} in Ref.~\onlinecite{Das2017}, who studied the non-local injection and detection of pure spin currents by Ni\textsubscript{80}Fe\textsubscript{20} that was deposited directly on a YIG film, showing a similar effect to the one observed for CoFeB here.
In this previous work, however, the two magnetic layers are in direct contact (exchange coupled) such that, regarding the spin signal modulation, both magnonic and electronic effects may be probed\cite{Cramer2018}.
In contrast, here the Cu interlayer ensures that purely electronic spin currents enter the CoFeB, suppressing potential magnonic contributions.
While this already may explain the larger amplitude change of the spin injection/detection efficiency observed for Ni\textsubscript{80}Fe\textsubscript{20} as compared to the CoFeB film studied here, due to the complexity of the systems further factors can have significant impact.
For example, the spin mixing conductance of the YIG/Ni\textsubscript{80}Fe\textsubscript{20} interface, which anyway might be larger than the one of YIG/Cu\cite{Kikuchi2015}, could also depend on the relative alignment of the YIG and Ni\textsubscript{80}Fe\textsubscript{20} magnetization due to varying spin reflectance and transmission coefficients\cite{Brataas2000}.
In addition, one has to consider spin memory loss\cite{Belashchenko2016} at the Cu/CoFeB interface and, certainly, the different intrinsic properties of the metallic ferromagnets\cite{Tsukahara2014,Zhang2015}.

Note that Das \textit{et al.} ascribe the magnetization orientation-dependent conversion of spin and charge information in metallic ferromagnets to a spin accumulation induced by the anomalous Hall effect\cite{Nagaosa2010} (AHE), denoting it \textit{anomalous spin Hall effect}.\cite{Das2017}
Reviewing the fundamentals of the AHE, one notes that the term \textit{anomalous} is used in a narrower sense to emphasize the distinct behavior and especially origin of the AHE as compared to the ordinary Hall effect.\cite{Nagaosa2010}
In terms of AHE and SHE, then again, both effects are based on the same fundamental mechanisms of spin-charge conversion\cite{Sinova2015}.
Now while spin-up and spin-down electrons in normal metals exhibit equal properties, majority and minority spin electrons in ferromagnets occupy exchange split bands with different densities of states and wave characters (e.g. \textit{s,p,d}) at the Fermi level\cite{Papaconstantopoulos2015}.
Majority and minority electron dynamics thus exhibit different spin-orbit potentials, yielding different spin-charge interconversion efficiencies and, on that account, we can expect a spin-dependent (inverse) spin Hall effect.

We finally turn to the comparison of the spin dependent SHE and ISHE.
We find that there is a difference in the amplitude modulation for the spin current detection $\left[\SI[separate-uncertainty=true]{17.2 \pm 3.4}{\percent}\right]$ (Fig.~\ref{fig:cofeb5}b) and generation $\left[\SI[separate-uncertainty=true]{5.3 \pm 2.6}{\percent}\right]$ (Fig.~\ref{fig:cofeb5}c), which may be due to the growth of a Cu/CoFeB/Ru multilayer stack.
While in both cases a spin current flows perpendicularly to the Cu/CoFeB interface, a large in-plane charge current flows through the wire when employed as injector.
Finite element simulations (see Supporting Information) show that spatially varying current densities in the multilayer, which are due to different resistivities of the single Cu, CoFeB and Ru\cite{haynes2014crc,seemann2011origin}) layers, lead to Oersted fields that can affect the SHE induced spin current.
The simulation data reveals that in the Cu layer significant out-of-plane Oersted field components are created near the wire edges, while the Oersted field in the CoFeB layer is insignificant in comparison.
Since the SHE effect creates spin currents with an in-plane spin polarization $\bm{\upmu}_\mathrm{s}$ at the Cu/CoFeB interface, this out-of-plane field may alter their amplitude due to a torque $\bm{\uptau} \propto \bm{\upmu}_\mathrm{s} \times \bm{H}_\mathrm{Oe}$.
However, to verify this mechanism, further studies beyond the scope of this work including varied multilayer configurations are required.

To summarize, the magnetization-orientation dependent generation and detection of spin currents in the metallic ferromagnet Co\textsubscript{60}Fe\textsubscript{20}B\textsubscript{20} has been probed in a non-local spin transport experiment.
Depending on the angle between the spin current polarization and the CoFeB magnetization direction, the spin-charge interconversion efficiency in CoFeB significantly changes.
In contrast to previous work comprising either Co\cite{Cramer2018} or Ni\textsubscript{80}Fe\textsubscript{20}\cite{Das2017} spin detectors, where due to the used sample structures both magnonic and electronic effects in the ferromagnetic detector may play a role, direct magnonic contributions by exchange coupling are suppressed here by the insertion of a non-magnetic Cu layer between CoFeB and the system propagating the spin current (YIG). 
This ensures that purely electronic spin currents are emitted or absorbed by the CoFeB, the consequence being that the electronic origin of the aforementioned signal modulation is exclusively studied and quantified, namely the spin-dependent (inverse) spin Hall effect.
Our results demonstrate the possibility to use CoFeB as an efficient spin current injector or detector for magnonic applications, with its spin-dependent (I)SHE offering an additional degree of freedom to tune the spin-charge interconversion in flexible novel magnonic devices.

We kindly acknowledge support by the Deutsche Forschungsgemeinschaft (DFG) (SPP 1538 Spin Caloric Transport, SFB TRR173 SPIN+X in Mainz), the Graduate School of Excellence Materials Science in Mainz (DFG/GSC 266), and the EU project INSPIN (FP7-ICT-2013-X 612759).
R.L. acknowledges the European Union’s Horizon 2020 research and innovation programme under the Marie Skłodowska-Curie grant agreement FAST number 752195.

\bibliography{bibliography}

\end{document}